\documentclass[prc,showpacs,twocolumn]{revtex4}
\usepackage{amsmath}
\usepackage{graphicx}
\def\pt{$p_T$ }
\def\dis{distribution}

\def\bq{\begin{equation}}
\def\eq{\end{equation}}
\def\pxy{$\psi(x, y)$}
\def\ps{\phi_s}

\begin{document}

\title 
{Dependence of Ridge Formation on Trigger Azimuth:  Correlated Emission Model}
\author   
{Charles B.\ Chiu$^1$ and  Rudolph C. Hwa$^{2}$}
\affiliation
 {$^1$Center for Particles and Fields and Department of Physics,
University of Texas at Austin, Austin, TX 78712, USA\\
$^2$Institute of Theoretical Science and Department of
Physics, University of Oregon, Eugene, OR 97403-5203, USA}

\begin{abstract} 
 
 Ridge formation in near-side correlation in heavy-ion collisions is studied in the framework of a phenomenological model, called Correlated Emission Model (CEM). Successive soft emissions due to jet-medium interaction lead to the enhancement of thermal partons  which follow the local flow directions. The correlation between the flow direction and the semihard parton direction is the major factor that causes the ridge formation to depend on the trigger direction relative to the reaction plane. With the use of a few parameters we have been able to reproduce the data on the ridge yields as functions of the trigger azimuthal angle for different centralities. An inside-outside asymmetry function is proposed to further probe the characteristics of the azimuthal correlation function. Insights are provided for the understanding of some detailed aspects of the centrality dependence.
\end{abstract}
 
\maketitle
\section{Introduction}

Experimental studies of dihadron correlations in heavy-ion collisions have revealed important information about the interaction of jets with dense medium \cite{ca,ja,aa}.  The observation of ridges on the near-side of triggers \cite{jp} has further stimulated a large number of theoretical investigations on the dynamical origin of ridge formation \cite{sv}-\cite{ch2}.  More detailed properties of the ridges are needed to discriminate the different models.  Recently, Feng has reported on the dependence of the ridge yield on the azimuthal angle $\phi_s$ between the trigger angle $\phi_T$ and the reaction plane $\Psi_{RP}$ \cite{af}.  At mid-central (20 - 60\%) Au-Au collisions the ridge yield is found to decrease rapidly with increasing $\phi_s$, a feature that has not been hinted by previous experiments, nor predicted theoretically.  The aim of this paper is to reproduce that feature in the framework of a new model which we call Correlated Emission Model (CEM).

Various properties of the ridges are already known experimentally.  The ridge yield increases significantly with $N_{\rm{part}}$, but decreases only slightly with trigger momentum $p^{\rm{trig}}_T$ \cite{jp}.  The dependence on the associated particle momentum $p^{\rm{assoc}}_T$ is exponential, its slope being nearly independent of $p^{\rm{trig}}_T$ \cite{jp}.  The baryon/meson ($B/M$) ratio in the ridge is large, comparable to that in the inclusive distributions \cite{cs,jb}.  Since the above properties are mostly revealed at intermediate $p_T$ ($< 6$ GeV/c) where the analyses have been carried out, those ridges are strongly correlated to semihard jets \cite{brief}.  Because of energy loss of the semihard parton in traversing the dense medium, most of those jets are created near the surface.  The lost energy enhances the thermal partons, which give rise to the ridge that is exponential in $p^{\rm{assoc}}_T$.  The large $B/M$ ratio in the ridge suggests that the hadronization process is recombination \cite{vg,rf,hy}.  When averaged over all trigger angle $\phi_T$, the $\Delta \phi$ distribution of the ridge is confined to the range $|\Delta \phi| <1$.  All these properties can be incorporated in a suitable model.  The challenge now is to reproduce the $\phi_s$ dependence of the ridge yield as found in \cite{af}.

It should be mentioned that there is another piece of data on the length of the ridge in $\Delta \eta$ that is new and striking.  PHOBOS has found that it extends over the range $-4 < \Delta \eta < 2$ \cite{ew}.  The model that we describe in this paper aimed at treating the $\phi_s$ dependence in the transverse plane is not suitable for describing the  $\Delta \eta$ dependence in the longitudinal direction for the same reason that correlations at large $p_T$ and large $p_L$ involves different physics.   We envision that the long-range $\Delta \eta$ correlation is due to the interaction between high-$k_T$ hard partons and the large-$k_L$ bulk partons, a subject that is not our concern in this paper.   Henceforth, we shall consider only correlation in the transverse plane with $|\eta|<1$, as it is restricted in the analysis in \cite{af}.

The range of $p_T$ studied for the $\phi_s$ dependence is limited to $3 < p^{\rm{trig}}_T < 4$ GeV/c and $1.5 < p^{\rm{assoc}}_T < 2.0$ GeV/c \cite{af}.  In fact, the experimental evidence for the exponential behavior of the ridge particles does not extend to $p^{\rm{assoc}}_T < 2 $ GeV/c, there being no data points on the $p^{\rm{assoc}}_T$ distribution below that threshold \cite{jp}.  Thus we shall not investigate the $p_T$ dependence in our model; to do so would involve issues, such as the recombination of thermal partons, that are not critical to our focus on the azimuthal problem.

We shall assume that the trigger angle $\phi _T$ is the angle of the semihard parton that initiates the trigger particle.  Event-by-event the two angles may not be identical, but on average that is not a bad approximation.  It is important to recognize that there is a significant difference in time scales between the dynamics of semihard partons and the formation of ridge particles.  The latter takes place at late time and is influenced by collective flow.  Compared to that time scale semihard scattering occurs almost instantaneously and is sensitive to the initial configuration of the collision system.  The difference in the time scales also leads to a difference in the azimuthal angles of the semihard parton $\phi _s$ and of a ridge particle $\phi$.  Our concern will mainly be in the correlation between $\phi_s$ and $\phi$ in the transverse plane, with parton momentum $k_T$ and hadron momentum $p_T$ essentially fixed in narrow ranges appropriate for the experimental $p^{\rm{trig}}_T$ and $p^{\rm{assoc}}_T$.  

Since the effects of semihard scattering cannot be calculated reliably from first principles, we build a model that incorporates all the features mentioned above concerning ridge formation.  The main characteristic about the model is the relationship between the local flow of the medium and the semihard parton that traverses that medium.  The interaction between the two leads to a correlation that increases the effect of soft emission when the enhanced thermal partons are all within a cone around the jet direction.  If the energy loss to the medium due to successive soft emission is restricted to the vicinity of the semihard parton in general agreement with the direction of collective flow, the enhancement of thermal partons that leads to ridge formation is clearly more effective than if the flow direction is normal to the jet direction.  We shall show that this correlation in the jet and flow directions will not only reproduce the $\phi_s$ dependence of the ridge yield, but also lead to other consequences that can readily be tested by analyzing available data in appropriate ways.  One such measure is the inside-outside asymmetry function whose properties we can predict.  Another result is the surprising feature that at fixed small $\phi_s$ the ridge yield per trigger has a bump as a function of impact parameter, i.e., it does not decrease monotonically with increasing peripherality .

The focus of our study in this paper is on the correlation between trigger and ridge particles on the near side.  The structure of the away-side distribution involves additional physics not relevant on the near side because of trigger bias that results in higher jet momentum on the away side in order to penetrate the bulk medium, leading to such phenomenon as the formation of double peak.  Nevertheless, the physics of ridge formation considered here may well be pertinent to the properties of the away-side peaks, a subject worthy of dedicated investigation in a separate study.

\section{Geometrical Considerations of Semihard Partons}

Since the trigger momentum is in the range $3 < p^{\rm{trig}}_T < 4$ GeV/c, the semihard partons are at least also in that range.  They lose some momentum in traversing medium, but gain some in recombination with thermal partons to form hadrons.  With parton $k_T > 3$ GeV/c, the time involved in its creation is roughly $k^{-1}_T < 0.1$ fm/c, which is short enough to be sensitive to the initial configuration of the system.  Let the point of creation in the almond-shaped overlap region be labeled by $\sf P$ with coordinates $(x_0, y_0)$ in the transverse plane, the $x$-axis being in the reaction plane, i.e.\ $\Psi _{RP} = 0$, and the $y$-axis being along the long side of the almond.  The semihard parton can be scattered into any azimuthal angle, which may differ from the trigger angle $\phi_T$ in any event, but on average they may be taken to be the same.  Thus hereafter we shall assume that the semihard parton's angle is $\phi_s$, the same as measured in the experiment \cite{af}, i.e.\ $\phi_s = \phi_T - \Psi _{RP}$.

Because of energy loss to the medium a semihard parton that can trigger an event is not likely to originate from the deep interior of the system.    Neither can it be right at the surface if it is to generate a ridge, since it has to interact with the medium and lose enough energy to enhance the thermal partons.  If $\sf P$ is a point at, say, 1 fm away from the surface, then in a time interval of 1 fm/c for the parton to reach the surface, the medium will have expanded and the geometry of the system may be better described by an expanding ellipse.  There is no reliable way to treat the problem, since hydrodynamics is not applicable without strong assumptions about fast thermalization at time $\tau < 1$
 fm/c, and QCD is not perturbative for semihard scattering.  In the model that we construct to describe the process, we consider a straight-line trajectory for the parton from $\sf P$ toward the boundary at angle $\phi_s$, ignoring the recoil parton that moves toward the interior and is absorbed by the bulk medium.  The boundary is to be described by an ellipse.  The distance between $\sf P$ and the boundary measured along the trajectory is to be denoted by $t$ (not time).  At points along the trajectory soft emission occurs that leads to the development of the ridge particles, the distribution of which will be discussed in the next section.  For now, let us focus on the geometry related to the trajectory of the semihard parton.
 
At impact parameter $b$ the initial system is almond shaped with width and height being
\begin{eqnarray}
w = R_A - b/2 \ , \quad  h = \left[R^2_A - (b/2)^2 \right]^{1/2} \ ,
\label{1}
\end{eqnarray}
where $R_A$ is the nuclear radius.  In the following we shall use dimensionless length variables by normalizing all lengths by $R_A$, so (\ref{1}) is to appear as
\begin{eqnarray}
w = 1 - b/2 \ , \quad  h = \left(1 - b^2/4 \right)^{1/2} \ .
\label{2}
\end{eqnarray}
An ellipse evolving from such a system is to be described by the equation
\begin{eqnarray}
\left({x \over w} \right)^2 +  \left({y \over h} \right)^2  = u \ .
\label{3}
\end{eqnarray}
The initial configuration corresponds to $u = 1$, so we use $(x_1, y_1)$ to denote the coordinates of that ellipse.  Since a semihard parton created at $\sf P$ is at a distance $t$ on the straight line at angle $\phi_s$ from the boundary, there is a short time interval for the parton to go that distance to reach the boundary.  We ignore the small expansion that the system may undergo during the transit time, since $\sf P$ is not far from the boundary.  The time when hadronization occurs is much later, when the medium density is lower, and $u$ larger.  Compared to that time scale, the system is almost static, while the parton traverses the medium.  This static approximation renders the determination of $t$ much easier.

The point at which the trajectory intersects the boundary is given by
\begin{eqnarray}
x_1 = x_0 + t \cos \phi_s \ , \quad y_1 = y_0 + t \sin \phi_s \ .
\label{4}
\end{eqnarray}
Using Eq.\ (\ref{3}) for $(x_1, y_1)$ at $u = 1$, we can solve for $t$, getting
\begin{eqnarray}
t = \left[\left(B^2 + AC \right)^{1/2} - B \right]/A \ ,
\label{5}
\end{eqnarray}
where
\begin{eqnarray}
A = \left({1\over w} \cos \phi_s \right)^2 +  \left({1\over h} \sin  \phi_s \right)^2 \ ,
\label{6}
\end{eqnarray}
\begin{eqnarray}
B = {x_0\over w^2} \cos \phi_s+ {y_0\over h^2} \sin  \phi_s  \ , 
\label{7}
\end{eqnarray}
\begin{eqnarray}
C = 1 - \left(x_0/w \right)^2  - \left( y_0/h\right)^2 \ .
\label{8}
\end{eqnarray}

At any point $(x, y)$ the local flow direction is specified by the gradient of $u(x, y)$ even for $u < 1$.  The azimuthal angle of that flow direction will be denoted by $\psi(x, y)$, whose value is
\begin{eqnarray}
\psi(x, y) = \tan ^{-1} \left(w^2y \over h^2x\right) \  .
\label{9}
\end{eqnarray}
Since for any $b$ the angle $\psi(x,y)$ can vary from $-\pi/2$ to $+\pi/2$ depending on the position of the point $(x,y)$ with $x>0$, there can always be a semihard-parton trajectory with a $\phi_s$ that coincides with $\psi(x,y)$. But it also means that there is a broad range of possibilities where $\phi_s$ differs from $\psi$. To emphasize this difference is the main characteristic of this model.
 
In the following section we shall have dynamical reason to follow the flow direction starting from any point $(x, y)$ on the parton trajectory along $\phi_s$.  Let $t'$ denote the distance from $(x, y)$ to the surface in the direction $\psi(x, y)$; it can be calculated in the same way as in Eqs.\ (\ref{4})-(\ref{8}) and the equation for $t'$ is the same as in Eq.\ (\ref{5}) except for the replacement of $\phi_s$ by $\psi(x, y)$.

We shall take the local density $D(x, y)$ in the transverse plane to be as described in the Glauter model for AB collision
\begin{eqnarray}
g_{AB}\left(\vec{b},  \vec{s} \right) = T_A (s) \left[ 1 - e ^{-\sigma T_B\left(\left| \vec{s} - \vec{b} \right| \right)}\right] \nonumber \\
+ T_B \left(\left| \vec{s} - \vec{b} \right| \right)  \left[ 1 - e ^{-\sigma T_A(s)}\right]
\label{10}
\end{eqnarray}
where $T_A (s)$ is the thickness function normalized to $A$, i.e.
\begin{eqnarray}
T_A (s) = A \int dz\ \rho (s, z)\ ,\quad  \int d^2s\ T_A(s) = A \ .
\label{11}
\end{eqnarray}
$\rho$ is the nuclear density normalized to $1$, for which we adopt the Woods-Saxon form
\begin{eqnarray}
\rho(r) = \rho_0 \left[ 1+e^{(r-r_0)/\xi}\right]^{-1} \ .
\label{12}
\end{eqnarray}
where $r_0=6.45$ fm and $\xi=0.55$ fm. We shall take the effective nuclear radius to be $R_A=7$ fm, which is also used to scale all length variables, so that when scaled the corresponding variables are $r_0=0.92$ and $\xi=0.08$. The scaled mean density $\rho_0$ is then 0.285, and $\sigma$ being the inelastic nucleon-nucleon cross section (taken to be 40 mb) becomes 0.082. For a point $(x, y)$ in the right-half almond region we have 
 \begin{eqnarray}
s^2 = (x + b/2)^2 + y^2 \ ,
\label{13}
\end{eqnarray}
so the longitudinal lengths of $A$ and $B$ at that point are
\begin{eqnarray}
L_{A,B} (x, y) = {1\over \rho_0}\int_{-z_{A,B}}^{z_{A,B}} dz \rho(s,z) \ ,
\label{14}
\end{eqnarray}
where 
 \begin{eqnarray}
z_A^2=1-s^2,  \qquad z_B^2=1-|\vec s-\vec b|^2 \ .
\label{15}
\end{eqnarray}
Relating Eq.\ (\ref{11}) to (\ref{14}) we have
 \begin{eqnarray}
\sigma T_A(s)=\omega L_A(x,y) ,  \qquad \omega={\sigma A \rho_0\over R_A^2}=4.6 \ ,
\label{15a}
\end{eqnarray}
where $A=197$ has been used. Thus apart from an overall normalization constant, the local density in the transverse plane is
\begin{eqnarray}
D (x, y) =  L_A (x, y) \left[1 - e^{-\omega L_B(x, y)}  \right]   \nonumber\\
+  L_B (x, y) \left[ 1 - e^{-\omega L_A(x, y)} \right]  \ .
\label{16}
\end{eqnarray}
The application of $D (x, y)$  below will not rely on its absolute magnitude.

\section{Ridge Formation}

Having described the geometry related to the semihard parton, we proceed now to the consideration of hadronization and ridge formation.  Since there is no theoretical framework in which one can reliably treat how the semihard parton interact with the medium and how the energy loss is converted to ridge particles, we propose a model that describes the subprocesses in terms of relevant distributions with parameters to be determined phenomenologically.    The conversion of lost energy to ridge particles is described by a correlation function, which turns out to be central to the phenomenology of $\phi _s$ dependence.

Let a semihard scattering occur at $\sf P$ in the transverse plane with a scattered parton moving at angle $\phi _s$.  The distance from $\sf P$ to the boundary along the straight-line trajectory is $t$.  
The probability $P(x_0,y_0,t)$ of detecting a parton emerging from the medium is the product of the probability of producing a semihard parton at $(x_0,y_0)$, which is proportional to the product of the longitudinal lengths at that point, $L_A(x_0,y_0)L_B(x_0,y_0)$, and the survival probability $S(t)$, i.e.,
\begin{eqnarray}
P(x_0,y_0,t)\propto L_A(x_0,y_0)L_B(x_0,y_0)S(t) \ .  \label{16a}
\end{eqnarray}
The proportionality factor that depends on the semihard scattering cross section will be canceled when we calculate the per-trigger yield, so it is not important to have it specified here. We assume that $S(t)$ has an exponential dependence on $t$ due to the opaqueness of the dense medium
\begin{eqnarray}
S(t)=\exp [-t/\tau(x_0,y_0)] \ ,  \label{17}
\end{eqnarray}
where $\tau(x_0,y_0)$ should  depend on the density along the trajectory. To implement the calculation in a manageable way, we first evaluate the density function $D(x_0,y_0)$ at all grid points within the geometric region of interest, and then between any given initial point $\sf P$ and the exit point we evaluate the local density along the corresponding $t$-segment by means of  2D interpolation among neighboring grid points. 
 Thus we write
\begin{eqnarray}
\tau(x_0,y_0)=t_0/d(x_0,y_0) \ ,  \label{17a}
\end{eqnarray}
where $t_0$ is a free parameter and $d(x_0,y_0)$ is the relative density
\begin{eqnarray}
d(x_0,y_0)=D(x_0,y_0)/D(0,0)  \label{17b}
\end{eqnarray}
with $D(0,0)$ being the density at the center of the overlap $x=y=0$. Since only semihard partons created near the surface are likely to lead to a trigger particle, $t_0$ is expected to be small, so we shall use just one such parameter for all trajectories at all centralities. Thus,
 if $t$ is large compared to $\tau(x_0,y_0)$, the parton would be absorbed by the medium, and cannot leave it to form a trigger particle or any structure above the bulk background.  In the numerical computation we cut off $t$ at $2\tau(x_0,y_0)$.
 
 To initiate a description of the soft interactions that generate the ridge, let  us first use
  $\xi t$ to denote the distance from $\sf P$ along that trajectory so that  $\xi = 1$ is at the boundary.  With $(x_0, y_0)$ being the coordinates of  $\sf P$, the coordinates $(x_{\xi}, y_{\xi})$ at $\xi$ are
\begin{eqnarray}
x_{\xi} = x_0 + \xi t \cos \phi_s \ , \quad  \ y_{\xi} = y_0 + \xi t \sin \phi_s  \ .
\label{18}
\end{eqnarray}
The probability that the semihard parton emits a soft parton at $\xi$ is proportional to 
 $D (x_{\xi}, y_{\xi})$.  We cannot specify in more detail the nature of the soft emission in the absence of a quantitative description of the soft process.  We assume at the qualitative level that some gluons are emitted by the semihard parton that do not significantly alter the straight-line trajectory of the parton.  Such soft emissions occur at successive points along the path.  If they augment one another coherently, a significant effect may accumulate and lead to observable consequences on hadronization.
 
Usual perturbative theory applied to the study of energy loss of hard partons traversing dense medium is not concerned with what happens to the medium.  Our concern here is the opposite.  The gluons radiated by a semihard parton are absorbed by the medium, thereby enhancing the thermal motion of the medium partons in the vicinity of the trajectory.  Since the medium expands, those thermal partons flow collectively and carry the extra energy gained along the flow whose direction can be determined locally.  We are interested in the thermal partons  because of our focus on ridge particles, which have the empirical characeristics that 
\begin{description}
\item[(a)]their $p^{\rm{assoc}}_T$ distribution is exponential with slope harder than that of the bulk background,
\item[(b)] their yield increases with centrality, and 
\item[(c)]the $B/M$ ratio in the ridge is similar to that in the inclusive distribution.  
\end{description}
All three combined strongly suggest  that the ridge particles are formed by recombination of enhanced thermal partons \cite{brief}.   The semihard parton that emerges from the medium leads to the creation of the trigger particle, but plays no direct role in the formation of the ridge.  The above discussion refers to the average over all triggered events.

We now direct our attention to the dependence on $\phi _s$.  In the discussion above two directions are emphasized:  one is the azimuthal angle $\phi _s$ of the trajectory; the other is the flow direction, denoted by $\psi$ in the preceding section. The former refers to the semihard parton, while the latter refers to the 
movement of the local medium that carries the 
soft partons in a direction that may or may not differ from $\ps$.  If $\psi(x,y)$ is approximately equal to $\phi_s$ for most of the points $(x,y)$ along the trajectory of the semihard parton, then the thermal partons enhanced by successive soft emissions are carried by the flow along in the same direction; the effects reinforce one another and lead to the formation of a ridge in a narrow cone. On the other hand, if the two directions are orthogonal, then the soft partons  emitted from the various points along the trajectory are dispersed over a range of surface area, so their hadronization leads to no pronounced effect. These  extreme possibilities suggest a correlation function between $\phi_s$ and $\psi$, which we assume to have the Gaussian form
\begin{eqnarray}
C(x, y, \phi_s) = \exp \left[ -{(\phi_s-\psi(x, y))^2\over 2 \lambda}\right] \ ,  
\label{19}
\end{eqnarray}
where the width-squared $\lambda$ is a parameter to be determined. This is a phenomenological formula that cannot be derived from first principles, but has sound physical basis and will play a central role in our model.

\begin{figure}[tbph]
\hspace{-.3cm}
\includegraphics[width=0.45\textwidth]{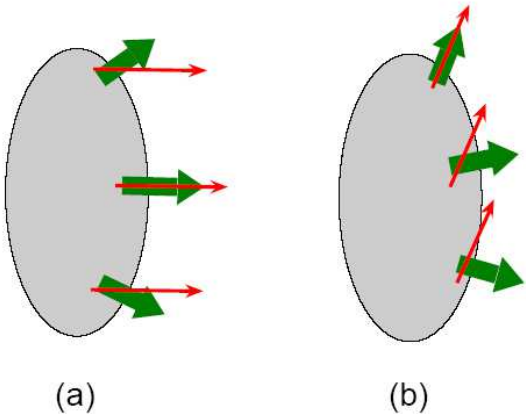}
\caption{ (Color online) Illustrations of the relationship between the trigger directions $\phi_s$ in (red) arrows and the flow directions $\psi$ in thick (green) arrows for noncentral collision. (a) Semihard partons at $\phi_s=0$ originated from 3 different points $\sf P$ where only the middle one has matching $\phi_s$ and $\psi$ that lead to strong ridge, while in (b) for $\phi_s\sim 70^{\circ}$ only the upper one has matching angles, leading to stronger ridge than in the two lower non-matching cases, but it is weaker than the middle one in (a) because of lower local density at the tip of the ellipse.}
\end{figure}

To provide a pictorial impression of the correlation between $\ps$ and $\psi$ that can affect the ridge formation, we show in Fig.\ 1 some examples of possible values of those angles. It is important to recognize that the point {\sf P} of creation of the semihard parton can vary over all points in the ellipse and that the possible mismatch between those angles depends on {\sf P}. In both panels of Fig.\ 1 we show three illustrative points of {\sf P}. In (a) we set $\ps=0^{\circ}$ shown by the thin arrows in red, and in (b) $\ps \sim 70^{\circ}$. The thick arrows indicate the flow directions $\psi$ normal to the surface. Evidently, the middle {\sf P} in (a) and the upper one in (b) result in matching $\ps$ and $\psi$, while the others do not. Ridge formation is stronger for the matching pairs than the mismatched ones. The ridge in (b) is weaker than that in (a) because the density near the top of the ellipse is lower than that in the middle. In essence, the origin of the $\ps$ dependence of the ridge is depicted in Fig.\ 1.

For every point $(x, y)$ on the trajectory, the flow direction $\psi(x, y)$ specifies only the average direction of the ridge hadrons, since there are statistical fluctuations, the magnitude of which depends on how far $(x, y)$ is away from the surface along the direction \pxy. That distance is $t'$, noted already in the preceding section. We introduce another \dis\ to describe the fluctuation of the azimuthal angle $\phi$ of a ridge particle from the average flow direction
\begin{eqnarray}
\Gamma(x,y,\phi) = \exp \left[ -{\left(\phi-\psi(x,y)\right)^2\over 2\gamma t'}\right] \ , 
\label{20}
\end{eqnarray}
where the degree of fluctuation is specified by $\gamma t'$; $t'$ is the height of the cone of fluctuation and $\gamma$ specifies the width. Clearly, the farther the emission point is away from the surface, the wider $\phi$ fluctuates from \pxy.
 
 We now assemble all the pieces that describe the various subprocesses and write the ridge-particle distribution $R(\phi,\phi_s,x_0,y_0)$ due to a semihard parton created at $(x_0,y_0)$ moving at $\phi_s$ as a product of all the factors, integrated along the trajectory:
\begin{eqnarray}
R(\phi,\phi_s,x_0,y_0) = N P(x_0,y_0,t) t \qquad\qquad \qquad \nonumber \\ 
\times \int_0^1 d\xi D(x_{\xi},y_{\xi}) 
C(x_{\xi},y_{\xi},\phi_s) \Gamma(x_{\xi},y_{\xi},\phi) \ ,  
 \label{21}
\end{eqnarray}
 where $N$ is an overall normalization constant. The variables $t,x_{\xi}$ and $y_{\xi}$ all depend implicitly on the initial coordinates $(x_0,y_0)$. For the observed \dis\ it is necessary to integrate over all $(x_0,y_0)$. Not every semihard parton included in that integration gets out of the medium to generate a particle that triggers the event. The ridge \dis\ per trigger is therefore normalized by the probability of the ridge-generating parton emerging from the medium
\begin{eqnarray}
 R(\phi, \phi_s) = {\int dx_0 dy_0 R(\phi,\phi_s,x_0,y_0)\over \int dx_0 dy_0 P(x_0,y_0,t)} \ . 
 \label{22}
 \end{eqnarray}
 In experimental analysis of the data positive and negative values of $\ps$ are combined to increase the statistics in the determination of $dN/d\Delta\phi$. We shall find, however, in our calculations interesting details of $R(\phi,\ps)$ that depend on the sign of $\ps$, as will be discussed below.
 
 In Eq.\ (\ref{22}) the integration of $(x_0,y_0)$ is over the entire initial ellipse specified by Eq.\ (\ref{3}) with $u=1$. For $-\pi/2<\ps<\pi/2$, the contribution comes mainly from the right-half ellipse, but not entirely.
At large $|y_0|$ and $x_0<0$ it is possible for a semihard parton to be emitted with $0<|\ps|<\pi/2$ and survive to trigger an event. Thus  we shall integrate over both positive and negative values of $x_0$ and $y_0$.

Equations  (\ref{21}) and (\ref{22}) contain the essential elements that affect the ridge distribution so long as the dependence on $p^{\rm{trig}}_T$ and $p^{\rm{assoc}}_T$ is not brought into the open.  Such dependencies enter into the parameters contained in the various factors, especially $S(t)$ and $C(x_{\xi}, y_{\xi}, \phi_s)$;  however, since we shall not  vary the ranges of $p^{\rm{trig}}_T$ and $p^{\rm{assoc}}_T$ in this paper, but just fix them at the narrow range of experimental values $3 < p^{\rm{trig}}_T < 4$ GeV/c and $1.5 < p^{\rm{assoc}}_T < 2.0$ GeV/c, we leave the $p_T$ dependence implicit, and proceed to the confrontation with real data.  

\section{$\phi_s$ Dependence of Ridge Yield}
 
Apart from the overall normalization $N$ in Eq.\ (\ref{21}), there are three essential parameters in our model:  $t_0$ in $S(t)$, $\lambda$ in $C(x, y, \phi_s)$, and $\gamma$ in $\Gamma (x, y, \phi)$.  They quantify three independent characteristics of the ridge formation process:  survivability, azimuthal correlation, and fluctuation, respectively. 

 Since the shape of the near-side correlation is insensitive to $\phi_s$, the integrated yield for $|\Delta\phi|<1$ has been analyzed for the ridge part and shown as a function of $\phi_s$ for $|\Delta \eta| < 0.7$ and for two centrality bins \cite{af}.  Thus we define accordingly
\begin{eqnarray}
Y(\phi_s) = \int^{\phi_s + 1}_{\phi_s - 1} d\phi\ R(\phi, \phi_s)
\label{23}
\end{eqnarray}
as the ridge yield per trigger for any impact parameter $b$. Carrying out the above integration over $R(\phi,\phi_s,x_0,y_0)$, where according to Eq.\ (\ref{21}) only the factor $\Gamma(x_{\xi},y_{\xi},\phi)$ depends on $\phi$, we obtain
\begin{eqnarray}
 \int^{\phi_s + 1}_{\phi_s - 1} d\phi\ 
R(\phi,\phi_s,x_0,y_0) =NP(x_0,y_0,t) t  \qquad\qquad \qquad \nonumber \\ 
\times\int_0^1 d\xi \ D(x_{\xi},y_{\xi}) C(x_{\xi},y_{\xi},\phi_s) G(x_{\xi},y_{\xi},\phi_s),\qquad  \label{24}
\end{eqnarray}
where
\bq
G(x_{\xi},y_{\xi},\phi_s) = \int_{\phi_s-1}^{\ps+1}d\phi\ \exp \left[-{\left(\phi-\psi(x_{\xi},y_{\xi})\right)^2\over 2\gamma t'}\right] .   \label{25}
\eq
The last integral is not sensitive to $\xi, \ps$ or $\gamma$, so $G(x_{\xi},y_{\xi},\phi_s)$ acts as a modifier of the overall normalization $N$. The parameter $\gamma$ will be determined in the next section by the differential correlation that depends on $\phi$, but here we focus first on the integrated yield, using  $\gamma=1$ that will be shown to be the final value. The point is that the yield $Y(\ps)$ depends mainly on two parameters: $t_0$ and $\lambda$.

  The yield data are shown in Fig.\ 2 for (a) top 5\% and (b) 20-60\%  with $\phi_s$ in 6 segments ranging from $0$ to $90^{\circ}$.  The mild dependence on $\ps$ in (a) is not surprising for central collisions, but the precipitous decrease  in (b) is striking.  The three main features of Fig.\ 2, namely, the relative normalization between the yields of the two centralities, the different rates of decrease with $\phi_s$, and the flattening out at large $\phi_s$, are correlated, and we fit them by varying the two parameters $t_0$ and $\lambda$.  The yield for the top 5\% is calculated by taking the average of $b=0.3$ and 0.4. 
The yield for the mid-central  case  is  obtained by averaging over  $b = 1$  and $1.3$, corresponding to the wider experimental  range of centrality 20-60\%. The solid lines in Fig.\ 2 represent the best fit we can achieve with values 
\begin{eqnarray}
t_0 = 0.2, \qquad  \lambda = 0.11\ .
\label{26}
\end{eqnarray}
In Fig.\ 2(a) the height of the solid line is adjusted to fit the data point at the lowest  $\ps$ by varying the normalization factor $N$ in Eq.\ (\ref{21}).  
$N$ encapsules all the uncalculable effects of the soft processes  involved in the ridge formation, and is not essential to the study of the $\ps$ dependence. However, once it is fixed at $N=0.085$ by the top 5\% data, the normalization for the mid-central collisions in Fig.\ 2(b) is no longer adjustable.  Our results reproduce the characteristics of the data very well for both central and mid-central collisions, but especially for 0-5\% centrality.  It is nontrivial
that  the decrease with increasing $\ps$ can be so different for the two centrality cases  and then they both flatten out above 60$^{\circ}$.

\begin{figure}[tbph]
\includegraphics[width=0.5\textwidth]{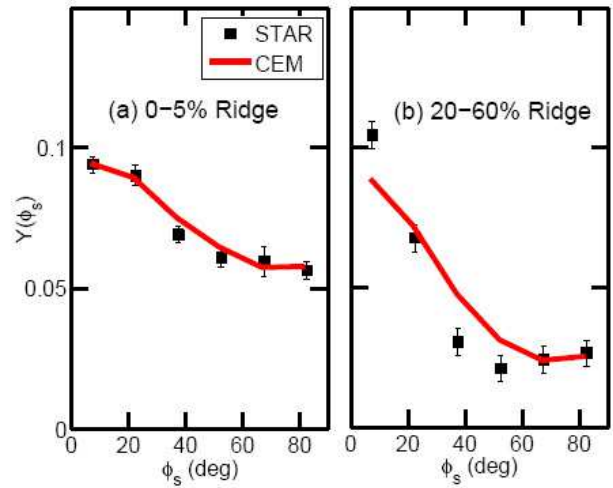}
\caption{(Color online) Dependence of ridge yield on $\ps$ for (a) top 5\% and (b) 20-60\%. Data are from Ref.\ \cite{af}. The solid lines are the results of calculation in CEM.}
\end{figure}

The values of the parameters in Eq.\ (\ref{26}) are very reasonable.  First, $t_0 = 0.2$ implies that only semihard partons created on the average at $R_A/4$ from the surface or closer get out to form trigger and ridge particles.  Second, $\lambda = 0.11$ implies that the width $\sigma_c= \sqrt{\lambda}$ of the correlation function is $0.34$ rad, a value that is significantly smaller, as it should, than $\Delta \phi \sim 1$, where the wings of the ridge vanish.  
The most important piece of physics that the phenomenology has revealed to us is contained in the correlation function $C(x, y, \phi_s)$ given in Eq.\ (\ref{19}).  It implies that
a strong ridge can be formed only if the local  flow direction $\psi (x, y)$ is within a cone of about $20^{\circ}$ from
 the direction of the semihard parton $\phi_s$.  The rapid decrease of the yield with $\phi_s$ seen in Fig.\ 2(b) is a direct consequence of the mismatch between the two directions.    The degree of that mismatch depends on the initial point $(x_0,y_0)$ for any fixed $\phi_s$.  As depicted in Fig.\ 1, if $\phi_s \approx 0$, the mismatch is larger at large $|y_0|$ than at small $|y_0|$, whereas if $\phi_s \approx \pi/2$, the opposite is true.  In the former case matching condition occurs (i.e., when $\ps\approx\psi\approx 0$) at small $|y_0|$,  where the density is high, thus enhancing soft emission; in the latter case (when $\ps\approx\psi\approx \pi/2$) it occurs at large $y_0$, where the density is low, thus suppressing ridge formation.  That is why the ridge yield decreases with increasing $\phi_s$, and does so more rapidly in noncentral collisions because the density varies more significantly near the surface of the overlap in that case. 
The leveling-off of the descent at high $\ps$ in both centrality cases is due to the contribution from the partons created in the left-half ellipse at high $y_0$. 

\section{$\Delta\phi$ Distribution of the Ridge }

Having reproduced the data on the $\ps$ dependence of the ridge yield, we now consider the $\Delta\phi$ dependence of the ridge particles. To that end we study the behavior of $R(\phi,\ps)$ in Eq.\ (\ref{22}) without integrating over $\phi$, as done in Eq.\ (\ref{23}). The data in Ref.\ \cite{af} show that the near-side peaks in $(1/N_{\rm trig}) dN/d\Delta\phi$, where $\Delta\phi=\phi-\ps$, have rather similar shape for different bins of $\ps$ and $p_T^{\rm assoc}$. 
For definiteness, we focus on the bin for $15^{\circ} < \ps < 30^{\circ}$ at  20-$60\%$ centrality, as shown in Fig.\ 3(a). They include both jet and ridge components. The dashed line is the calculated result for $R(\phi,\ps)$ at $\ps=22^{\circ}$ (left curve) and the dashed-dotted line is for $\ps=-22^{\circ}$ (right curve), plotted as functions of $\phi-\ps$. Both are calculated  for $\gamma=1$. Due to the symmetry in the problem the two curves are mirror reflections of each other about $\Delta\phi=0$. Since the data on the $\Delta\phi$ \dis\ include the contributions from both positive and negative values of $\ps$, we average the two curves and obtain the solid line for $|\ps|=22^{\circ}$. The difference between the data and the calculated $R(\Delta\phi,|\ps|)$ is the jet contribution, $J(\Delta\phi,|\ps|)$. In the visual presentation of Ref.\ \cite{af}, not included in the proceedings, the data for the ridge distribution are shown; the segment for $15^{\circ} < \ps < 30^{\circ}$ at  20-$60\%$ centrality is reproduced in Fig.\ 3(b). The solid line in that plot is our calculated result for $R(\phi,\ps)$ with $\gamma=1$ [identical to that in Fig.\ 3(a)], showing good agreement with the data. 

\begin{figure}[tbph]
\vspace{-.3cm}
\includegraphics[width=0.5\textwidth]{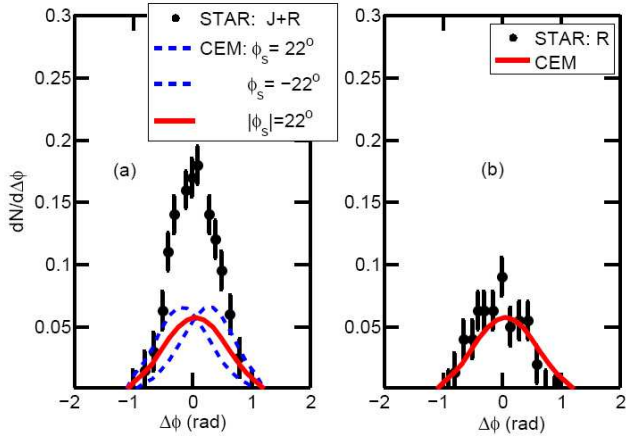}
\caption{(Color online) The data are $\Delta\phi$ distributions from \cite{af} for $15<\ps<30^{\circ}$ at 20-60\% centrality  for (a) the sum of jet and ridge and (b) ridge alone. The curves are all calculated in the CEM for the ridge \dis s only with $\gamma=1$. The dashed and dashed-dotted lines are left- and right-shifted for $\ps=\pm 22^{\circ}$, respectively. The solid lines are the average over the two signs of $\ps$.}
\end{figure}

To see the sensitivity of our result to the value of $\gamma$, we show in Fig.\ 4 two curves for $R(\Delta\phi,|\ps|)$ for $\gamma=1$ (solid) and $\gamma=2$ (dashed). In view of the large error bars both are acceptable, although the solid line agrees better with the data on both sides of $\Delta\phi=0$.
Hereafter, we shall regard $\gamma=1$ as the  value determined from fitting the correlation data.

\begin{figure}[tbph]
\includegraphics[width=0.45\textwidth]{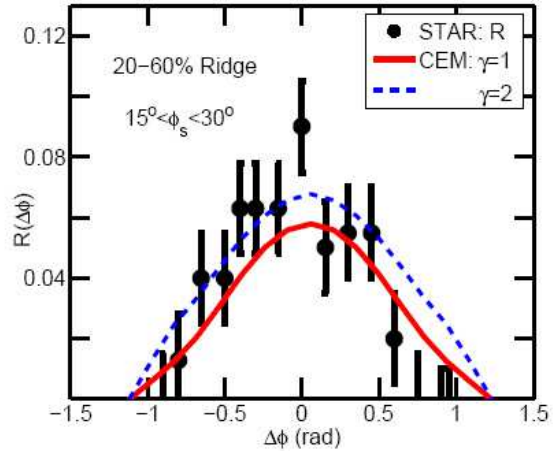}
\caption{(Color online) Same as Fig.\ 2(b) with the solid line for $\gamma=1$ and dashed line for $\gamma=2$.}
\end{figure}

The asymmetric result for  positive and negative $\ps$ shown in Fig.\ 3(a) is significant. For positive $\ps$, the shift of the peak of $R(\Delta\phi,\ps=22^{\circ})$ to the left of $\Delta\phi=0$ shown in Fig.\ 3(a) can be studied for other values of $\ps$ also. We show in Fig.\ 5 our calculated results for $R(\phi,\ps)$ at $\ps=7^{\circ}, 22^{\circ}, 37^{\circ}, 52^{\circ}, 67^{\circ}$ and $82^{\circ}$. The leftward displacements from $\Delta\phi=0$ reach their maximum at $\ps \approx 37^{\circ}$ where the magnitude of the shift is approximately $10^{\circ}$. We regard this asymmetry of the ridge \dis s as an indication of some important aspect of the physics, which we  shall in the next section quantify  in a  way  that may be easier for experiments to verify.

\begin{figure}[tbph]
\includegraphics[width=0.5\textwidth]{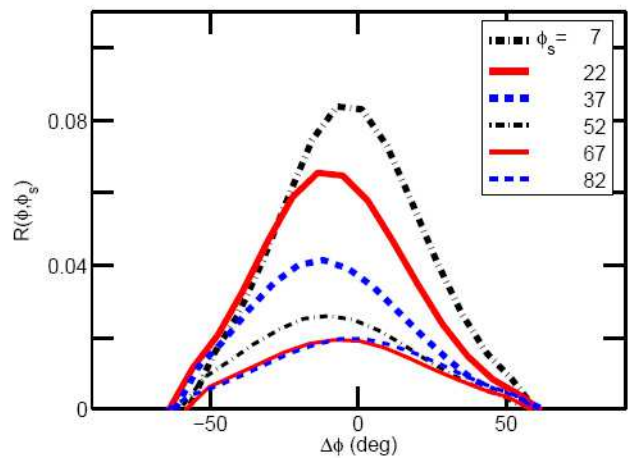}
\caption{(Color online) The ridge \dis s for various positive values of $\ps$.}
\end{figure}

\section{Inside-outside Asymmetry Function}

The shifting of the maxima of the ridge \dis s seen in Figs.\ 3(a) and 5 reveals some property of the correlation between semihard parton and the ridge particles, although the phenomenon is not seen directly in the correlation data that collect both positive and negative values of $\ps$. The ridge yields in Fig.\ 2 reveal only the gross effect of the correlation because of the integration over $\phi$ in Eq.\ (\ref{23}), but not some detailed properties.    To have a qualitative understanding of the origin of the shift,  let us consider the case $\phi_s = \pi/4$, for instance, bearing in mind the difference between $\phi_s$  and $\psi(x, y)$.  For most values of $(x, y)$ along the parton trajectories, the flow directions normal to the surface have $\psi(x, y) < \pi/4$.  However, for $(x_0, y_0)$ in a small region near the top of the ellipse, $\psi(x, y)$ is $> \pi/4$, but that is where the density is low. Thus in most of the regions where the semihard partons are produced with $\phi_s = \pi/4$, the ridge hadrons have $\phi$ directions mostly smaller than the parton direction. We therefore expect an asymmetry between $\phi < \phi_s$ and  $\phi > \phi_s$. To quantify these properties in a way accessible to experimental analysis, we now introduce a measure that we refer to as the inside-outside asymmetry.
 
 Restricting $\phi_s$ to the range $0 < \phi_s < \pi/2$ we define
\begin{subequations}
 \begin{eqnarray}
Y_+ (\phi_s) = \int^{\phi_s}_{\phi_s - 1} d\phi R(\phi, \phi_s), \label{27a}\\ 
 Y_- (\phi_s) = \int^{\phi_s+1}_{\phi_s} d\phi R(\phi, \phi_s) \ . 
\label{27b}
\end{eqnarray}
\end{subequations}
But for $-\pi/2 < \ps < 0$ we define them in the opposite way
\begin{subequations}
\begin{eqnarray}
Y_+ (\phi_s) = \int_{\phi_s}^{\phi_s+1} d\phi R(\phi, \phi_s),  \label{28a}   \\ 
Y_- (\phi_s) = \int^{\phi_s}_{\phi_s-1} d\phi R(\phi, \phi_s) \ . 
\label{28b}
\end{eqnarray}
\end{subequations}
Thus $Y_+$ may be thought of as being mostly inside (i.e., on the in-plane side of $\phi_T$), while $Y_-$ may be thought of as being mostly outside (i.e., on the out-of-plane side of $\phi_T$). Clearly, when $\ps=0$, it is necessary that $Y_+$ equals $Y_-$ to be consistent, and of course they are equal by reflection symmetry across the horizontal axis. Also, at $\ps=\pm\pi/2$ reflection symmetry across the vertical axis requires $Y_+=Y_-$.
We define the asymmetry function to be
\begin{eqnarray}
A (\phi_s) = {Y_+ (\phi_s)- Y_- (\phi_s) \over Y_+ (\phi_s) + Y_- (\phi_s) } \ , 
\label{29}
\end{eqnarray}
where the denominator is just $Y(\phi_s)$ as defined in Eq.\ (\ref{23}).  It follows from reflection symmetries that  $A(0) = 0$ and $A(\pm\pi/2) = 0$.  How $A(\phi_s)$ varies between the two extremes at $\ps=0$ and $\pi/2$ depends on  $b$. Experimentally, the two definitions of $Y_{\pm}(\ps)$ in Eqs.\ (27) and (28) are necessary in order that the data in the two quadrants of $\ps$ may be combined to increase statistics, but theoretically, one is the reflection of the other, so the study of the sector $0<\ps<\pi/2$ is sufficient.

\begin{figure}[tbph]
\includegraphics[width=0.45\textwidth]{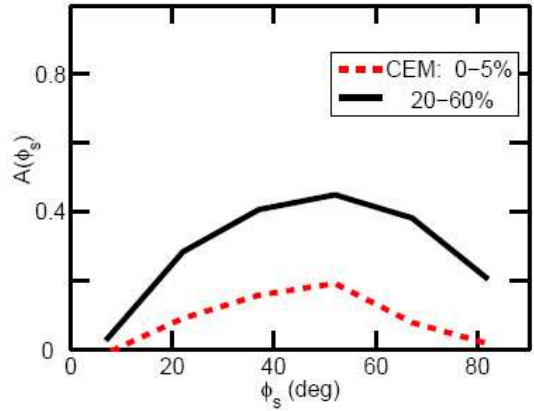}
\caption{(Color online) The asymmetry function $A(\ps)$ for 0-5\% (dashed) and 20-60\% (solid).}
\end{figure}

In Fig.\ 6 we show the asymmetry function for the two centrality bins. It is evident that for the entire range of $\ps$ the asymmetry $A(\ps)$ for mid-central collisions is more than twice  larger than that for central collisions. Of course, at $b=0$ there should be no asymmetry. The rapid growth of $A(\ps)$  with increasing $b$ is a striking feature of the effects of the mismatch between $\ps$ and $\psi$ that complements the $\ps$ dependence of the ridge yield. Experimental verification of  our prediction of the properties of $A(\ps)$ would lend further support to  the correlated emission model.

\section{Centrality Dependence}

We have investigated more thoroughly the centrality dependence of the yield per trigger, and show in Fig.\ 7(a) our result for $Y(b,\ps)$ versus impact parameter $b$ for  various values of $\ps$. The monotonic decrease of $Y(b,\ps)$ with increasing $b$ occurs only  for $\ps> 40^{\circ}$. At lower $\ps$ it increases with $b$ initially before decreasing. Such a peaking is an unexpected finding. However, after averaging over all $\ps$ the bump in the per-trigger yield disappears. For the averaging process we first note that the denominator of $R(\phi,\ps)$ defined in Eq.\ (\ref{22}) has implicit dependence on $\ps$ because $t$ given in Eq.\ (\ref{5}) does. Thus the average yield is obtained by averaging both the numerator and denominator of $Y(b,\ps)$ over $\ps$ separately, i.e.,
\begin{eqnarray}
Y(b)={\int_0^{\pi/2}d\ps \int_{\ps-1}^{\ps+1}d\phi \int dx_0dy_0 R(\phi,\ps,x_0,y_0)\over \int_0^{\pi/2}d\ps \int dx_0dy_0 P(x_0,y_0,t)},  \label{30}
\end{eqnarray}
where the dependencies on $b$ in the integrands are implicit. It is this $Y(b)$ that is plotted in Fig.\ 7(b).
When plotted against $N_{\rm part}$ the average yield increases monotonically, as shown in Fig.\ 7(c). Since the magnitude of the per-trigger yield depends sensitively on the cut in $p_T^{\rm assoc}$, we do not show in that figure any data having cuts different from $1.5<p_T^{\rm assoc}<2.0$ GeV/c. The two points in that figure are determined from the data in Ref.\ {\cite{af}}, using the number of triggers in each $\ps$ bin provided by the group that performed the analysis.

\begin{figure}[tbph]
\includegraphics[width=0.5\textwidth]{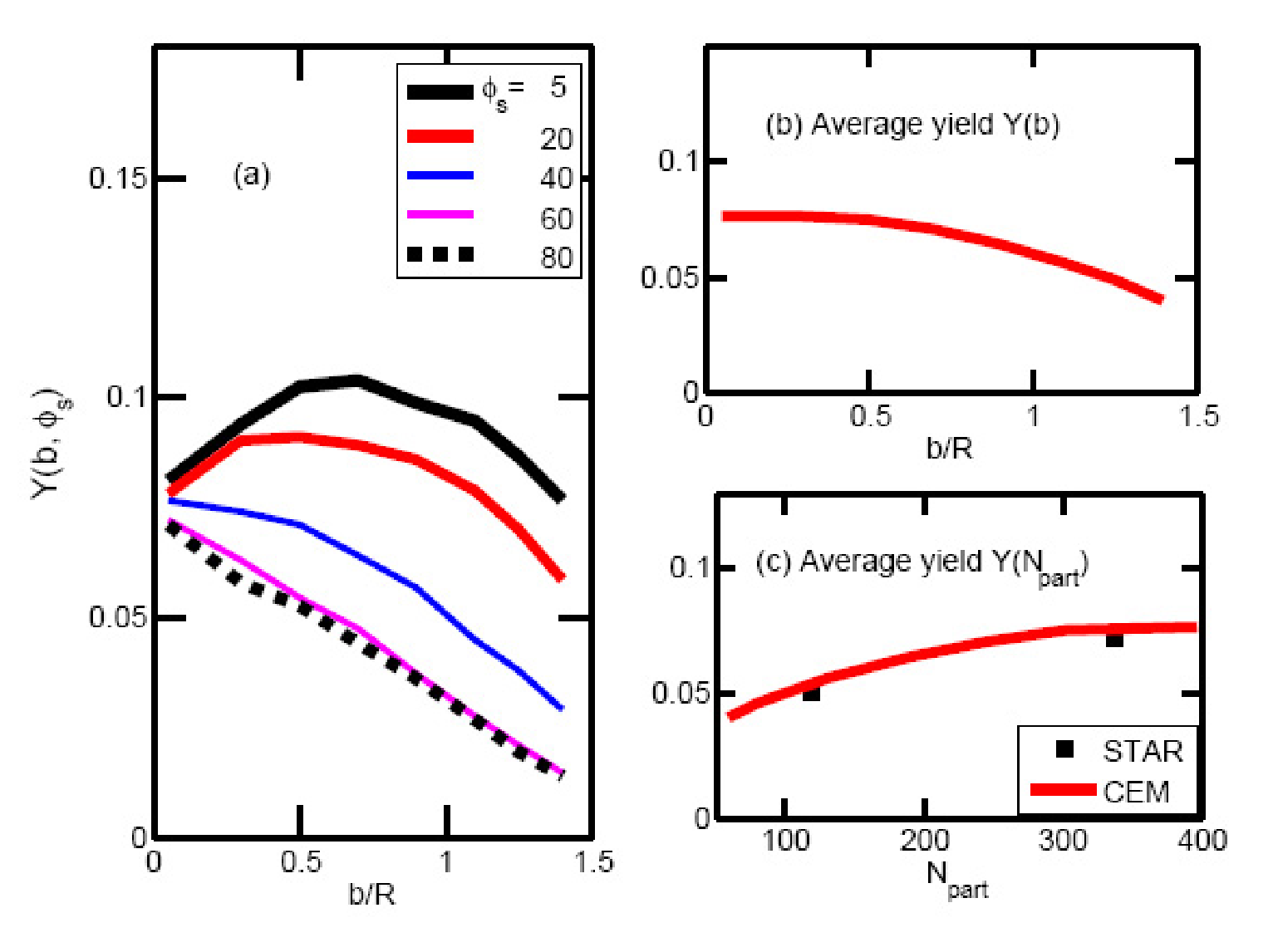}
\caption{(Color online) (a)  Ridge yield per trigger vs impact parameter for 5 values of $\ps$,  (b) $Y(b)$, averaged over all $\ps$, vs impact parameter, and (c) average yield vs $N_{\rm part}$. The two points in (c) are determined from the data in Fig.\ 2(a) and (b).}
\end{figure}

It is of interest to remark on the significance of the peak of $Y(b,\ps)$ at mid-$b$ for small $\ps$ in Fig.\ 7(a). The peak suggests that at small $\ps$ the yield can be higher in noncentral collisions than in central collisions. That is a theoretical result that has  phenomenological support as can be seen  in Fig.\ 2, where the data point at the lowest $\ps$ at $7^{\circ}$ in panel (b) is higher than that in panel (a).  Our calculated result in Fig.\ 2(b) for $Y(\ps)$, after integrating over the range of centrality 20-60\%, shows a slower descent with $\ps$ compared to the data, and at $\ps=7^{\circ}$ it is $\sim 10\%$ lower than the data point. It means that the peaking of our theoretical result on $Y(b,\ps)$ in Fig.\ 7(a), though remarkable in its existence, is not high enough. The most 
likely source of the discrepancy lies in the value of $\lambda$ for mid-values of $b$, since a decrease of $\lambda$ for $1<b<1.3$ can produce a steeper descend in $\ps$. However, we prefer at this stage of our investigation not to introduce extra freedom to achieve a better fit of the data. 
 In our view the dynamical origin of the observed phenomenon at low $\ps$ is more important than the numerical accuracy of our result. As we now proceed to describe, our model provides an explanation of the small $\ps$ behavior.

 The geometrical difference between mid-central and central collisions is, of course, the extra overlap space in the latter case.  In that extra space the deep interior does not make any contribution to the ridge, since the semihard partons originated there cannot survive on their way to the surface.  The only regions where the partons moving at small $\phi_s$ can get out are at the top and bottom of the ellipse; i.e., those partons form tangential jets.  However, the flow vectors in those regions are essentially normal to the parton trajectories.  Thus the correlation function $C (x, y, \phi_s)$ suppresses the formation of ridge particles for partons originating in the extra space.  Yet the number of trigger particles is not suppressed because of the tangential jets.  Consequently, the yield/trigger decreases with decreasing $b$.  Note that the argument does not work if $\phi_s$ is large because then $\psi (x, y)$ would be close to $\phi_s$ and there would be no suppression due to $C (x, y, \phi_s)$.

Support for our assertion that ridge formation by tangential jets is suppressed can be found in  3-particle correlation  in dijet events where the two trigger jets have nearly equal and opposite momenta. It has been reported that preliminary results from STAR indicate the lack of ridge structure in events with one trigger at $p_T>5$ GeV/c, another at $p_T>4$ GeV/c and associated particles at $p_T>1.5$ GeV/c in AuAu collisions at any centrality \cite{ob}. From the $N_{\rm part}^{2/3}$ dependence of the number of dijets triggered, it is suggested that those events are due to tangential jets generated near the surface. The analysis in Ref.\ \cite{ob} makes no reference to $\ps$ and integrates over all $\phi_T$. Clearly, tangential jets can be directed at any azimuthal angle for any $N_{\rm part}$. Our explanation of the phenomenon is that for tangential jets the flow direction is normal to the semihard parton direction at any  $\ps$. We predict that 
as the momentum difference of the two trigger jets increases, corresponding to their becoming less tangential, the mismatch between $\ps$ and $\psi$ becomes less severe, and a ridge should begin to develop.

A phenomenon related to the dip at small $b$ has been observed previously by PHENIX in the dependence of the yield on $N_{\rm part}$ for baryon triggered jets (without ridge separation) \cite{ada}. The data show a dip of the yield at high $N_{\rm part}$.
Such behavior of centrality dependence has never been explained before.  Our study here cannot address the issue of species dependence of the trigger, and the data of Ref.\ {\cite{ada}} cannot be used to address the issue of ridge yield. Nevertheless, our result in Fig.\ 7(a) holds the intriguing possibility that our description of ridge formation may contain the germ of an explanation of the dip phenomenon.

 \section{Conclusion}
 
 Much theoretical attention has been given to the modification of the jet behavior at very high $p_t$ due to the propagation of a hard parton through dense medium. The focus has been on what happens to that hard parton. Our interest in this paper has been on what happens to the medium when a semihard parton passes through it. It is the complement to the problem at high $p_t$. Both are important, since jet-medium interaction acts in both ways. The effect of semihard partons on the medium is difficult to calculate, so a phenomenological model is needed to relate different existing empirical facts and to generate a coherent picture of what the important subprocesses are.
 
 Ridge formation provides crucial evidence that the medium's response is correlated to the jet direction. Because of the time lag in hadron production, local hydrodynamical flow of the medium can influence the direction in which the lost energy goes and where the ridge is to be formed. We have attempted to capture these aspects of the dynamics by use of some phenomenological functions in the CEM. What we have learned is that ridge formation can be strong only if the direction of the flow that carries the enhanced thermal partons is within a narrow cone of about 20$^{\circ}$ around the semihard-parton direction. Within that cone the effects of successive soft emissions by the parton reinforce one another to develop a ridge. When a parton's trajectory is normal to the flow direction, the effect of energy loss is spread out over a wide spatial domain and the detection of any ridge formation is suppressed.
 
 We have shown that the CEM reproduces the $\ps$ dependence of the ridge yield. Significant insight has been gained by examining the details of the ridge characteristics. As a result, we have made predictions on certain properties of the ridges that can be measured. One is on the inside-outside asymmetry behavior relative to the trigger direction. Another is on the centrality dependence of the ridge yield. Still another is on the lack of ridge structure in dijet events. Verification of our predictions will lend additional support to our interpretation  of  the medium response to jets created in heavy-ion collisions. 
 
 Although hadronization by recombination has not been used explicitly in our calculation, it is a subprocess that is implied in our modeling. The conversion of the energy lost by the semihard parton to the enhancement of thermal partons cannot be described quantitatively, so the angular \dis\ of the hadrons formed by the recombination of the thermal partons cannot be determined in the absence of calculable \dis\ of the soft partons. The $\Delta\phi$ dependence of the ridge has been studied phenomenologically in Ref.\ \cite{ch}, when the dependence on $\ps$ is not an issue. Here we have gone beyond that first attempt and have focused on the $\ps$ dependence. Our use of the correlation function $C(x,y,\ps)$ is based on the identification of the average hadronic direction with the local flow direction $\psi(x,y)$ of the soft partons by virtue of thermal-thermal recombination. 
 One may argue that the identification of the average hadronic direction with the flow direction is also an attribute of other schemes of hadronization, such as Cooper-Frye and local parton-hadron duality. However, it is only in thermal-thermal recombination that the $B/M$ ratio can be as high as observed \cite{cs,brief}.
 If ridge formation were dominated by thermal-shower recombination, we would have had to calculate the shower parton \dis, as done in Ref.\ \cite{chy} for $v_2$ at intermediate $p_T$, and the hadron direction in the ridge would not be simply determined by the flow direction of the thermal partons. Clearly, if ridges were formed by means of fragmentation of hard or semihard partons, there would be no suppression due to a mismatch of parton and flow directions. As a consequence, there would not be any $\ps$ dependence of the ridge formation as observed in the experiment.
 
 We have not investigated in this paper the dependence on \pt for either the trigger or the ridge particles. Neither have we considered the hadron species of those particles. For all $p_T<6$ GeV/c recombination is the only viable mechanism for hadronization \cite{brief}. To relate the various dependencies (i.e., on $p_T, \phi, \ps, \eta, N_{\rm part}$, hadron species in trigger, ridge, and jet) is still a larger problem that remains to be studied.
 
\section*{Acknowledgment}
We are grateful to Fuqiang Wang for helpful discussions and communication. This work was supported, in part,  by the U.\ S.\ 
Department of Energy under Grant No. DE-FG02-92ER40972.

%\newpage


\begin{thebibliography}{99}

%1
\bibitem{ca}
C.\ Adler {\it et al.}, STAR Collaboration,  Phys.\
Rev.\ Lett.\  {\bf 90}, 082302 (2003).

%2
\bibitem{ja}
J.\ Adams {\it et al.} (STAR Collaboration), Phys.\ Rev.\ Lett.\ {\bf 
95}, 152301 (2005); {\bf 97}, 162301 (2006).

%3
\bibitem{aa}
A.\ Adare {\it et al.},  (PHENIX Collaboration), Phys.\ Rev.\ C {\bf 77}, 011901 (R) (2008).

%4
\bibitem{jp}
J.\ Putschke (for STAR Collaboration), J.\ Phys.\ G {\bf 34}, S679 (2007).

%5
\bibitem{sv}
S.\ A.\ Voloshin, Nucl.\ Phys.\ A {\bf 749}, 287 (2005).

%6
\bibitem{ch}
C.\ B.\ Chiu and R.\ C.\ Hwa, Phys.\ Rev.\ C {\bf 72}, 034903 (2005).

%7
\bibitem{na}
N.\ Armesto, C.\ A.\ Salgado and U.\ A.\ Wiedemann, Phys.\ Rev.\ Lett.\  {\bf 93}, 242301 (2004).

%8
\bibitem{pr}
P.\ Romatsche,  Phys.\ Rev.\ C {\bf 75}, 014901 (2007).

%9
\bibitem{am}
A.\ Majumder, B.\ M\"{u}ller and S.\ A.\ Bass,  Phys.\ Rev.\ Lett.\ {\bf 99}, 042301 (2007).

%10
\bibitem{es}
E.\ V.\ Shuryak,  Phys.\ Rev.\ C {\bf 76}, 047901 (2007).

%11
\bibitem{cw}
C.\ Y.\ Wong, Phys.\ Rev.\ C {\bf 76}, 054908 (2007).

%12
\bibitem{ch2}
C.\ B.\ Chiu and R.\ C.\ Hwa, Phys.\ Rev.\ C {\bf 76}, 024904 (2007).

%13
\bibitem{af}
A.\ Feng,  (for STAR Collaboration), talk given at Quark Matter 2008, Jaipur, India (2008), J.\ Phys.\ G: Nucl.\ Part.\ Phys.\ {\bf 35}, 104082 (2008), arXiv: 0807.4606.

%14
\bibitem{cs}
C.\ Suarez,  (for STAR Collaboration), poster at Quark Matter 2008, Jaipur, India (2008).

%15
\bibitem{jb}
J.\ Bielcikova (for STAR Collaboration), talk given at Winter Meeting on Nuclear Physics, Bormio, Italy (2008).

%16
\bibitem{brief}
For a brief review see, for example, R.\ C.\ Hwa, plenary talk given at Quark Matter 2008, Jaipur, India (2008), J.\ Phys.\ G: Nucl.\ Part.\ Phys.\ {\bf 35}, 104017 (2008), arXiv:  0804.3763.

%17
\bibitem{vg} V.\ Greco, C.\ M.\ Ko, and P.\ L\'{e}vai, Phys.\ Rev.\ C {\bf 68}, 034904 (2003).

%18
\bibitem{rf}
R.\ J.\ Fries, B. M\"{u}ller, C.\ Nonaka and S.\ A.\ Bass,   Phys.\ Rev.\ C {\bf 68}, 044902 (2003).

%19
\bibitem{hy}
R.\ C.\ Hwa and C.\ B.\  Yang,  Phys.\ Rev.\ C {\bf 70}, 024905 (2004).

%20
\bibitem{ew}
E.\ Wenger (for PHOBOS Collaboration), talk given at Quark Matter 2008, Jaipur, India (2008), J.\ Phys.\ G: Nucl.\ Part.\ Phys.\ {\bf 35}, 104080 (2008), arXiv:  0804.3038.

%21
\bibitem{ob}
O.\ Barannikova, talk given at the International Symposium on Multiparticle Dynamics 2007, Berkeley, CA (2007); talk given at Quark Matter 2008, Jaipur, India (2008), J.\ Phys.\ G: Nucl.\ Part.\ Phys.\ {\bf 35}, 104086 (2008).

%22
\bibitem{ada}
A.\ Adare {\it et al.} (PHENIX Collaboration), Phys. Lett. B {\bf 649}, 359 (2007).

%23
\bibitem{chy}
C.\ B.\ Chiu, R.\ C.\ Hwa, and C.\ B.\ Yang, Phys.\ Rev.\ C {\bf 78}, 044903 (2008), revised version of arXiv: 0801.2183.
 
\end{thebibliography}
\end{document}